\documentclass[aps,amssymb,showpacs,amsmath,superscriptaddress,showkeys,prl, twocolumn]{revtex4-1}
\usepackage[utf8x]{inputenc}
\usepackage{graphicx}
\usepackage{epsf}
\usepackage{natbib}
\usepackage{units}
\usepackage{ulem}

\usepackage{graphicx}
\usepackage{units}
\usepackage[mediumspace,mediumqspace,squaren]{SIunits}

\newcommand{\reffig}[1]{Fig.\ref{#1}}

\newcommand{\PP}[1]{probe particle }

\begin{document}

\title{The origin of high-resolution IETS-STM images of organic molecules with functionalized tips }

\author{Prokop Hapala}
\email[corresponding author: ]{hapala@fzu.cz}
\affiliation{Institute of Physics, Academy of Sciences of the Czech Republic, v.v.i.,  Cukrovarnick\' a 10, 162 00 Prague, Czech Republic}
\author{F. Stefan Tautz}
\affiliation{Peter Gr{\"u}nberg Institut (PGI-3),Forschungszentrum J{\"u}lich, 52425 J{\"u}lich, Germany}
\affiliation{ J{\"u}lich Aachen Research Alliance (JARA)--Fundamentals of Future Information Technology, 52425 J{\"u}lich, Germany}
\author{Ruslan Temirov}
\affiliation{Peter Gr{\"u}nberg Institut (PGI-3),Forschungszentrum J{\"u}lich, 52425 J{\"u}lich, Germany}
\affiliation{ J{\"u}lich Aachen Research Alliance (JARA)--Fundamentals of Future Information Technology, 52425 J{\"u}lich, Germany}
\author{Pavel Jel\'{i}nek}
\email{jelinkep@fzu.cz}
\affiliation{Institute of Physics, Academy of Sciences of the Czech Republic, v.v.i.,  Cukrovarnick\' a 10, 162 00 Prague, Czech Republic}
\affiliation{Graduate School of Engineering, Osaka University 2-1, Yamada-Oka, Suita, Osaka 565-0871, Japan}


\begin{abstract}

Recently, the family of high-resolution scanning probe imaging techniques using decorated tips has been complimented by a method based on inelastic electron tunneling spectroscopy (IETS). The new technique resolves the inner structure of organic molecules by mapping the vibrational energy of a single carbonmonoxide (CO) molecule positioned at the apex of a scanning tunnelling microscope  (STM) tip. Here, we explain high-resolution IETS imaging by extending the model developed earlier for STM and atomic force microscopy (AFM) imaging with decorated tips. In particular, we show that the tip decorated with CO acts as a nanoscale sensor that changes the energy of the CO frustrated translation in response to the change of the local curvature of the surface potential. In addition, we show that high resolution AFM, STM and IETS-STM images can  deliver information about intramolecular charge transfer for molecules deposited on a~surface. To demonstrate this, we extended our numerical model by taking into the account the electrostatic force acting between the decorated tip and surface Hartree potential.
\end{abstract}


\maketitle


%

One of the most exciting and significant breakthroughs in field of scanning probe microscopy (SPM) in last years is undoubtedly the achievement of high-resolution STM \cite{Temirov2008} and AFM \cite{Gross_Science09} images of molecular structures with functionalized tips \cite{Bartels_PRL98,Weiss2010a}. In general, the high-resolution images, being typically acquired in the regime where the tip-surface interaction becomes repulsive, are characterized by a~presence of sharp features in both intra and intermolecular regions. The sharp ridges observed in the intramolecular region often mimics the internal molecular structure \cite{Gross2012}, with only few exceptions \cite{Pavlicek2013, Hapala_PRB_2014}. The capability of AFM/STM to resolve internal atomic and chemical structure in real space opened new horizons for the characterization of molecules  and surfaces \cite{Gross2012,Kawai_ACSNano2013,Neu2014,Welker_2012,Sun_2011,Swart_2011,Gross_NatChem_2010,Riss_NanoLett2014,Boneschanscher2014} at atomic scale.   

The origin of high resolution of molecular structures in AFM has been attributed to Pauli repulsion  \cite{Gross_Science09,Moll_NJP10}  and the bending of the functionalized tip apex \cite{Gross2012}. Recently, we introduced a~numerical model \cite{Hapala_PRB_2014} which provides a unified insight into the detailed mechanism of the high resolution imaging with decorated tips in both AFM and STM. According to the model, the decorated tip apex acts as an~atomistic force sensor that responds with significant relaxations of the decorating particle (probe particle) towards local minima of the tip-sample interaction potential at close distances. The relaxations cause discontinuities in both the frequency shift and the tunneling current signals and thus become observable in AFM and STM images as sharp contrast features. Although the model was originally used to confirm the decisive role of Pauli repulsion in the high-resolution STM and AFM imaging of molecular structures done with functionalized tips, it must be noted that imaging of other types of interactions (e.g. electrostatic) should also be possible  \cite{Gross2012}. Namely, the influence of intramolecular charge transfer on the molecular contrast has not been analyzed in detail yet.

Inelastic electron tunneling spectroscopy (IETS) \cite{Jaklevic_PRL66} is a well established technique, which has been used to perform,  e.g.  chemical identification \cite{Stipe_Science98_ChI} or reaction \cite{Pascual_Nature03, Stipe_PRL97, Katano_Science07} and molecular manipulation \cite{Stipe_Science98_Man}. Very recently, Chiang et al. \cite{Chiang_Science2014} introduced a~novel approach of the high-resolution molecular imaging by means of IETS. They obtained the high resolution images of a Cobalto-Phtalocyaine (CoPc) molecule deposited on the metal surface by mapping the IETS feature corresponding to the frustrated translational vibration mode of a CO molecule attached to the STM tip. The IETS-STM maps of Chiang et al. show the sharply resolved structure of the molecular skeleton (i.e. position of atoms and bonds), very similar to the high-resolution AFM/STM images with functionalized tips. Also the constant height STM image (see Fig. S4 in  \cite{Chiang_Science2014}) recorded during IETS mapping shows the characteristic sharp contrast features similar to those observed in earlier STM experiments \cite{Temirov2008, Weiss2010a, Kichin2013}. Further proliferation of the IETS-STM imaging method as well as the precise interpretation of the experimental results strongly depends on a~detailed understanding of the imaging mechanism. However, the underlying mechanism of the high-resolution IETS-STM images has not been addressed yet. 

In this Letter, we provide an explanation of the origin of the high-resolution IETS-STM images \cite{Chiang_Science2014}.  We show how the frustrated translation mode of the probe particle terminating the tip varies according to the changes of the surface potential. The IETS-STM imaging can thus be seen as another utilization of the nanoscale sensor functionality of the decorated tip mapping the interactions between the tip and the surface and transducing it into the characteristic IETS signal. We also extend our numerical AFM/STM model including the electrostatic force acting between functionalized probe and surface Hartree potential. This allows us to analyze in detail the influence of the electrostatic force on the high-resolution images. By applying the extended model to the analysis of the IETS-STM images of CoPc/Ag(110) reported by Chiang et al. \cite{Chiang_Science2014}, we demonstrate that the decorated tip is also sensitive to the electrostatic interaction between a~charge localized on the tip apex and an~electrostatic polarization generated by intramolecular charge transfer inside the CoPc molecule. 

We start by briefly summarizing the basic ingredients of our mechanistic model \cite{Hapala_PRB_2014}. A single atom or a small molecule decorating the apex of a metal tip is modeled by a single point particle interacting via van der Waals attraction and Pauli repulsion with the atoms of the surface. These interactions are described by an empirical pairwise Lenard-Jones (L-J) potential. The repulsive branch of the L-J potential describes the Pauli repulsion, that plays the key role in the high resolution imaging. Additionally, movement of the probe particle is constrained by a~lateral harmonic potential under the metallic tip apex. The harmonic potential determines the lateral "bending" stiffness $k$ of the functionalized tip. Finally we note that the electrostatic interactions that may occur between the charge density distributions in the tip and the surface have not been taken into account in the original model. 

On top of the mechanical model we simulate the STM images by computing the tunneling current through the junction at each lateral position of the tip as a two step tunneling process between the tip and the probe particle and further between the probe particle and the sample. As it has been shown previously, our model reproduces the numerous features of experimental high-resolution AFM/STM images very well.  More details about the model and its comparison with experimental evidence can be found here \cite{Hapala_PRB_2014}. 

To address the IETS-STM imagining experiments \cite{Chiang_Science2014}, we have extended our model to include a~calculation of the vibrational energy levels $\varepsilon$ of the probe particle. In addition, we have included the electrostatic force, which describes the Coulombic interaction between the internal charge on a surface/molecule and a charge cloud localized on the probe particle. For the detailed description of the new model see \cite{SI}.  

\begin{figure}
\centering
\includegraphics[width=8.5cm]{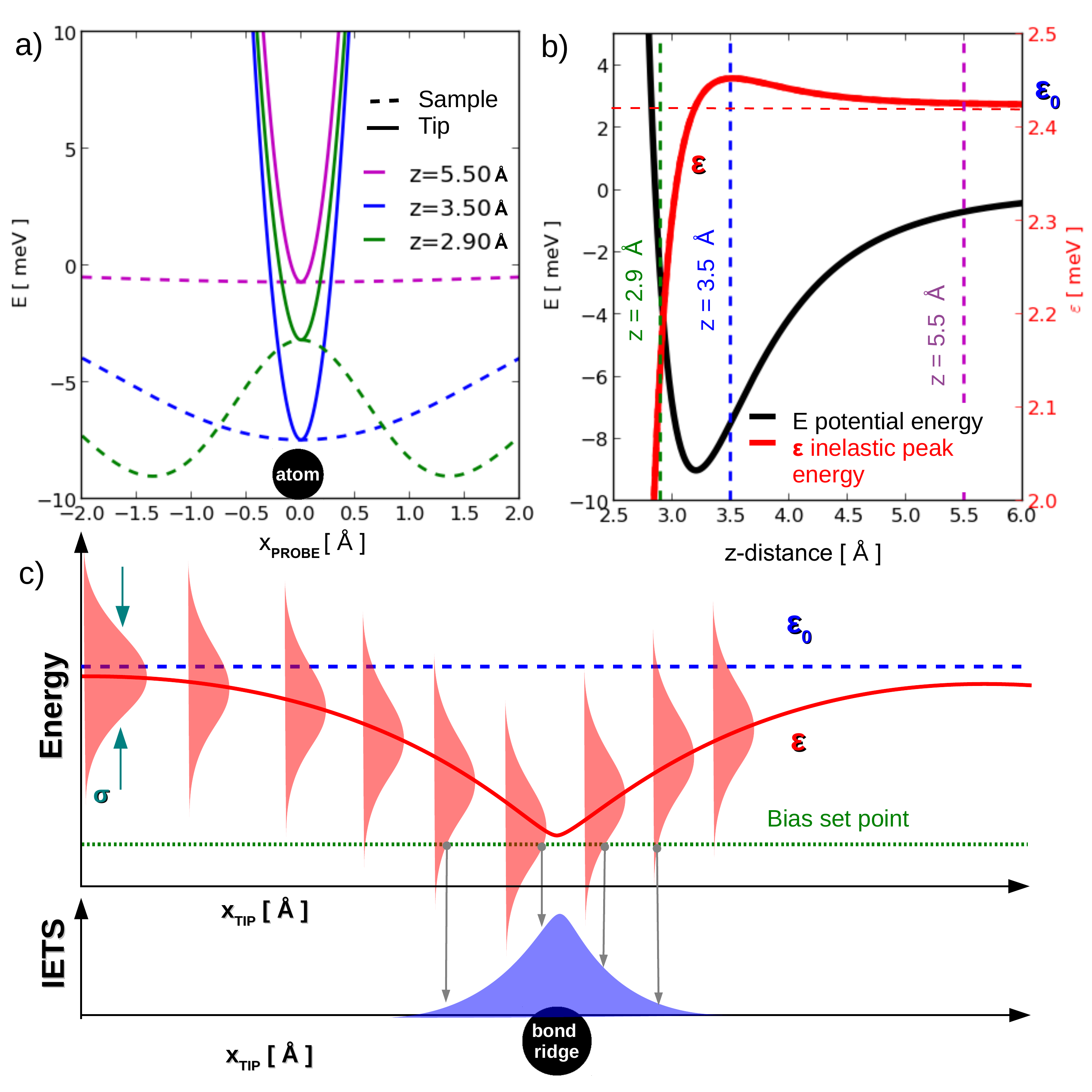}
\caption{\textbf{1D IETS-STM model} a) lateral view of the harmonic potential from the tip (full line) and Lenard-Jones potential from the surface atom (dashed line) for various distances $z$ of the probe particle over the surface atom. b) Corresponding evolution of the total potential energy $E$ of the probe particle (black line) and the lowest vibration energy eigenvalue $\varepsilon$ of the probe particle (red line) along the $z-$distance. c) Schematic explanation how the variation of the  vibration energy eigenvalue $\varepsilon$ (red line) affects the IETS-STM signal (blue peak) when the tip scans in $x-$direction over the bond ridge. In the repulsive regime, the proximity of the surface atom/bond ridge produces softening of the vibration energy eigenvalue $\varepsilon$  with respect to the free-standing vibration energy eigenvalue $\varepsilon_0$ (dashed blue line). Thus  a~characteristic IETS peak (represented by red gaussian) centered at the vibration energy $\varepsilon$ crosses 
a~bias set point (green line) at a certain $x$-distance accordingly changing the IETS-STM signal (blue peak).}
\label{fig-01}
\end{figure}

For didactic purposes, we first discuss a~simple system consisting of only a~single atom on a surface interacting with the \PP\. This allows us to identify the influence of the tip-sample interaction on the variation of the vibration mode of the \PP (e.g. CO molecule) and consequently the IETS signal. We found both the probe particle relaxations and variation of the vibrational mode energy $\varepsilon$ are driven by: (i) an~attractive potential of the STM-tip apex modeled as a harmonic spring (parabola); and (ii) van der Waals and Pauli repulsion potential of the surface atoms modeled by L-J potential (see \cite{SI}).  \reffig{fig-01}a presents the dependence of the tip and sample potentials acting on the \PP\ placed above the surface as a function of tip-sample distance. 

The red line in \reffig{fig-01}b shows the variation of the vibrational energy $\varepsilon$ along the z-distance over the surface atom. At the same time, the  total potential energy $E$ (black line in \reffig{fig-01}B) varies due to the interaction between the surface atom and the molecular probe. In the far distance regime, z=5.5 \AA, the interaction with the substrate is negligible. Therefore, the potential affecting the motion of the \PP\ is entirely determined by the lateral harmonic potential of the tip. In our simulations we set the the stiffness of this potential to  $k$=1.44 N/m. This provides the vibrational energy of the frustrated translation mode ~2.4 meV, which is in a good agreement with the experimental observation \cite{Chiang_Science2014}. The chosen $k$ corresponds well to the upper range predicted for CO-terminated tips (see supplementary of \cite{Gross2012}). A~relatively high stiffness value causes the lateral tip potential to dominate within the entire range of tip-sample distances. This potential is only slightly moderated by the contribution from the surface potential as the tip approaches closer to the sample. At closer tip-sample distance, z=3.5 \AA, the probe particle experiences the attractive surface potential as just a moderate perturbation. Nevertheless the presence of an attractive surface potential with convex curvature increases slightly ($< 0.1 meV$) the vibration energy $\varepsilon$. On the other hand, at very close distance, z=2.9 \AA,  the repulsive interaction of the surface potential has a strongly concave curvature near the maximum at the position of the surface atom (see dashed green line in \reffig{fig-01}a). This on one hand leads to the lateral relaxation of the probe particle and considerable softening $\approx 0.3 meV$ of its frustrated translational vibration mode $\varepsilon$ as well.

Let us generalize this observation for a 2D situation, where the tip scans in lateral direction over the atom at a close distance. In this case, the repulsion and hence the concave curvature of the surface potential is highest in areas directly above the molecular skeleton (i.e. local maxima and saddles).  \reffig{fig-01}c draws schematically the variation of the frustrated translation vibration energy $\varepsilon$ when the tip moves laterally with respect to the surface bond ridge or atom. We see that $\varepsilon$ and consequently the position of the IETS peak \cite{note1} decreases as the \PP\ approaches towards the bond, with a~sharp minimum just over the bond. Therefore lateral mapping of the IETS peak intensity made at a properly selected bias voltage (depicted by dashed green line in \reffig{fig-01}c) gives rise to the imaging contrast that exhibit significant variations over atoms/bonds. This explains the variation of the atomic contrast of IETS signal over the atoms and bond ridges observed in the experiment \cite{Chiang_Science2014}. We should stress, that in this analysis we do not take into account the \PP\ relaxation. As we will demonstrate later, the probe particle relaxation leads to the sharpening of the IETS-STM contrast as in the case of high-resolution AFM and STM imaging.

\begin{figure}
\centering
\includegraphics[width=8.5cm]{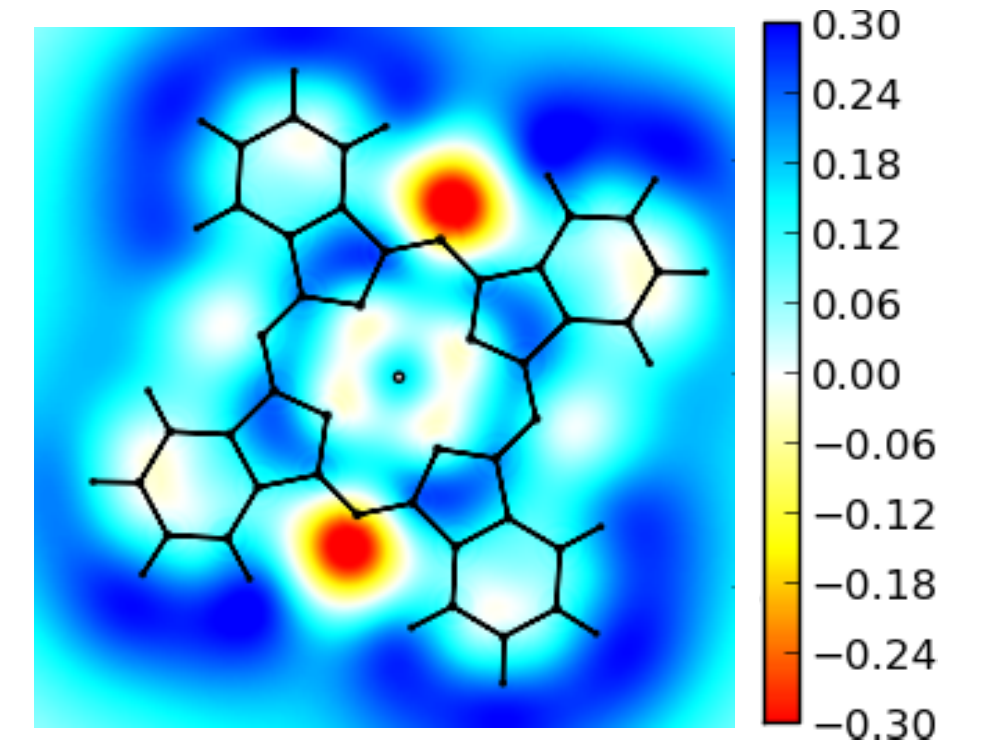}
\caption{\textbf{Hartree potential} of a CoPc molecule deposited on a Ag(110) surface obtained from DFT calculations in $xy$-plane cut 2.00 \AA\ above Co atom of the molecule. The Hartree potential reveals substantial intramolecular charge transfer between the pyrrole rings and imine nitrogens. Color bar scale is plotted in eV units.}
\label{fig-03}
\end{figure}

Now we benchmark our model against experimental results of a CoPc molecule deposited on a Ag(110) surface  \cite{Chiang_Science2014}. First, we carry out total energy DFT calculations with VASP code in order to obtain an optimized structure of the molecule on top of the surface. A detailed description of the DFT calculations can be found in \cite{SI}. According to the DFT calculations, the molecule is located $~\approx$ 3 \AA\ above the surface and the Co and N atom relaxes slightly downward, establishing a chemical bond with Ag underneath. On the other hand, the outer benzene rings move slightly upward so that the molecular corrugation is $~\approx$ 0.6 \AA\ . 

More importantly, the DFT simulation (see \cite{SI}) indicates  a~substantial charge transfer within the CoPc molecule. The Hartree potential shown in \reffig{fig-03} reveals positive potential over the pentagonal pyrrole ring and a negative potential over the imine nitrogens. In order to account for the possible effects caused by the charge transfer inside CoPc, we extend the original model \cite{Hapala_PRB_2014} by including the electrostatic interactions between the probe particle and the surface. The electrostatic force is evaluated from the Hartree surface potential obtained from DFT calculations and a~preselected  charge density on the probe particle (for details see \cite{SI}). As will be shown below, the inclusion of electrostatic interactions between the tip and the surface is vital for the understanding of the experimental IETS-STM contrast.

\begin{figure}
\centering
\includegraphics[width=8.5cm]{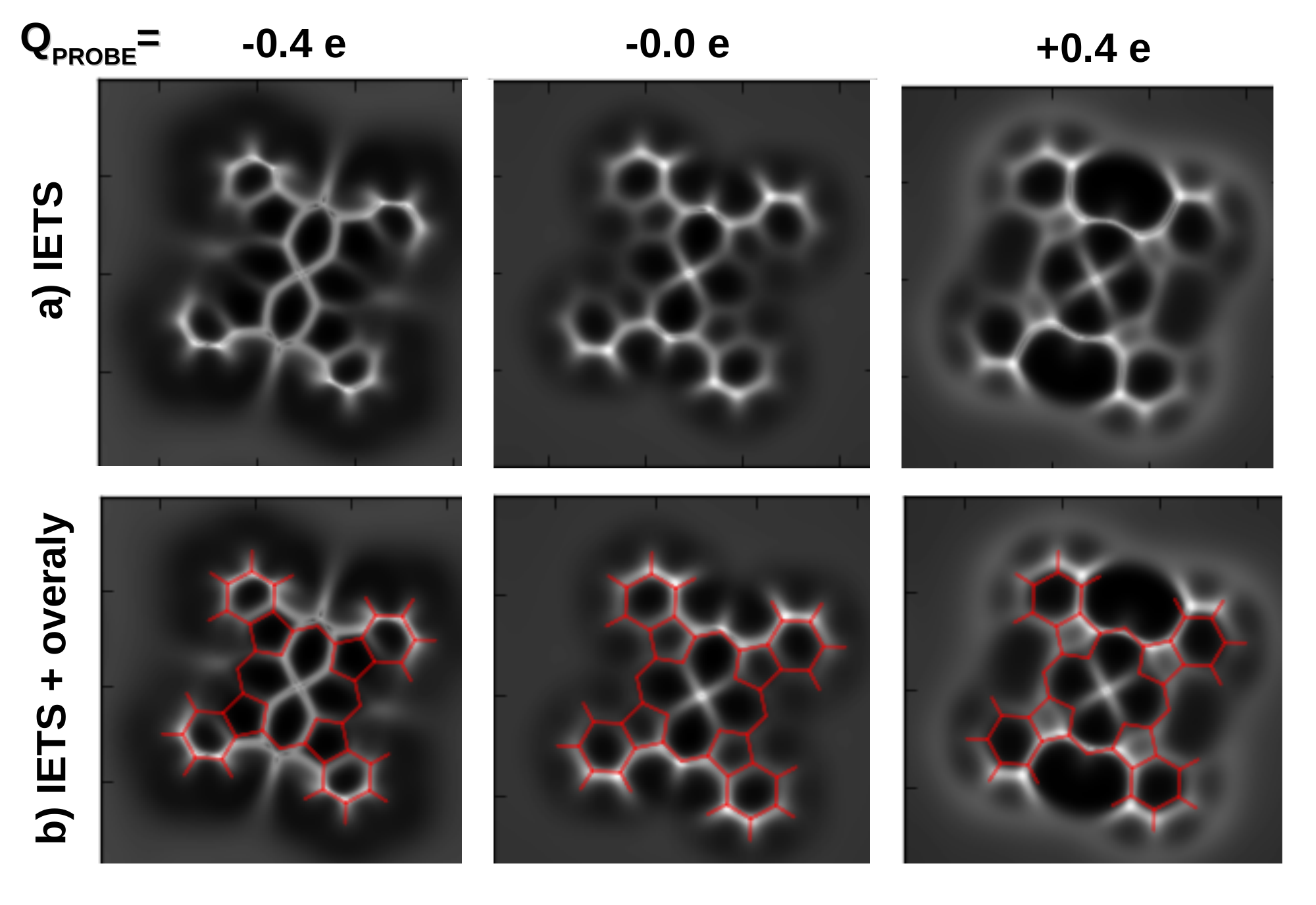}
\caption{\textbf{Simulated constant-height IETS-STM images} calculated in the distance $z$= 7.3 \AA\  at V=1.5 mV for different point charge values on \PP\ , Q=-0.4e, 0e and +0.4e. a) Bare IETS-STM images to visualize clearly all features b) The same images with overlaid  by molecular skeleton (red line) of the relaxed CoPc molecule on a Ag(110) surface obtained by the DFT calculations. Bright color means large intensity of the IETS peak following the scheme drawn on \reffig{fig-01}a. }
\label{fig-02}
\end{figure}

Detailed analysis of the experimental IETS-STM images of a CoPc molecule reveals an~enlargement of the pyrrole rings together with a~reduction of area corresponding to imine nitrogen in the central part of the molecule (see Fig. 3 in  \cite{Chiang_Science2014}). This is the area, where most of the internal charge transfer takes place, according to our analysis of the Hartree potential.  \reffig{fig-02} shows our simulated constant-height IETS-STM images obtained with the extended numerical model, including the electrostatic interactions for different values of charge located at the probe particle.  We see that the inclusion of the electrostatic interaction distorts the molecular contrast on the central part of the molecule (red line in \reffig{fig-02} depicting molecular skeleton of the relaxed molecule). The calculated IETS-STM image (\reffig{fig-02} a)) obtained with the negatively charged probe particle $Q$=-0.4e  matches very well the experimental evidence ( compare to Fig. 3A  in \cite{Chiang_Science2014}).  Alternatively, the presence of the positive charge $Q$=+0.4e on the \PP\  leads to an~opposite effect as visible on \reffig{fig-02}. 

We should note that our simulated constant-height IETS-STM images display a~more pronounced contrast on the outer benzene rings compared to the experiment \cite{Chiang_Science2014}. This discrepancy can be explained by a~large flexibility of the outer benzene rings. According to our DFT simulations, the benzene rings are only weakly coupled to the silver surface. Thus, when the tip operates in the repulsive regime, one can expect the flexible benzene rings of CoPc to bend down under the force exerted by the tip.  We assume the bending to be responsible for the reduced resolution over the benzene rings in the experiment. In contrast to the experiment, atomic structure of the molecules on surface is fixed rigidly. Consequently, this gives rise to the enhanced atomic contrast over the benzene rings in comparison with the experiment.

\begin{figure}
\centering
\includegraphics[width=8.5cm]{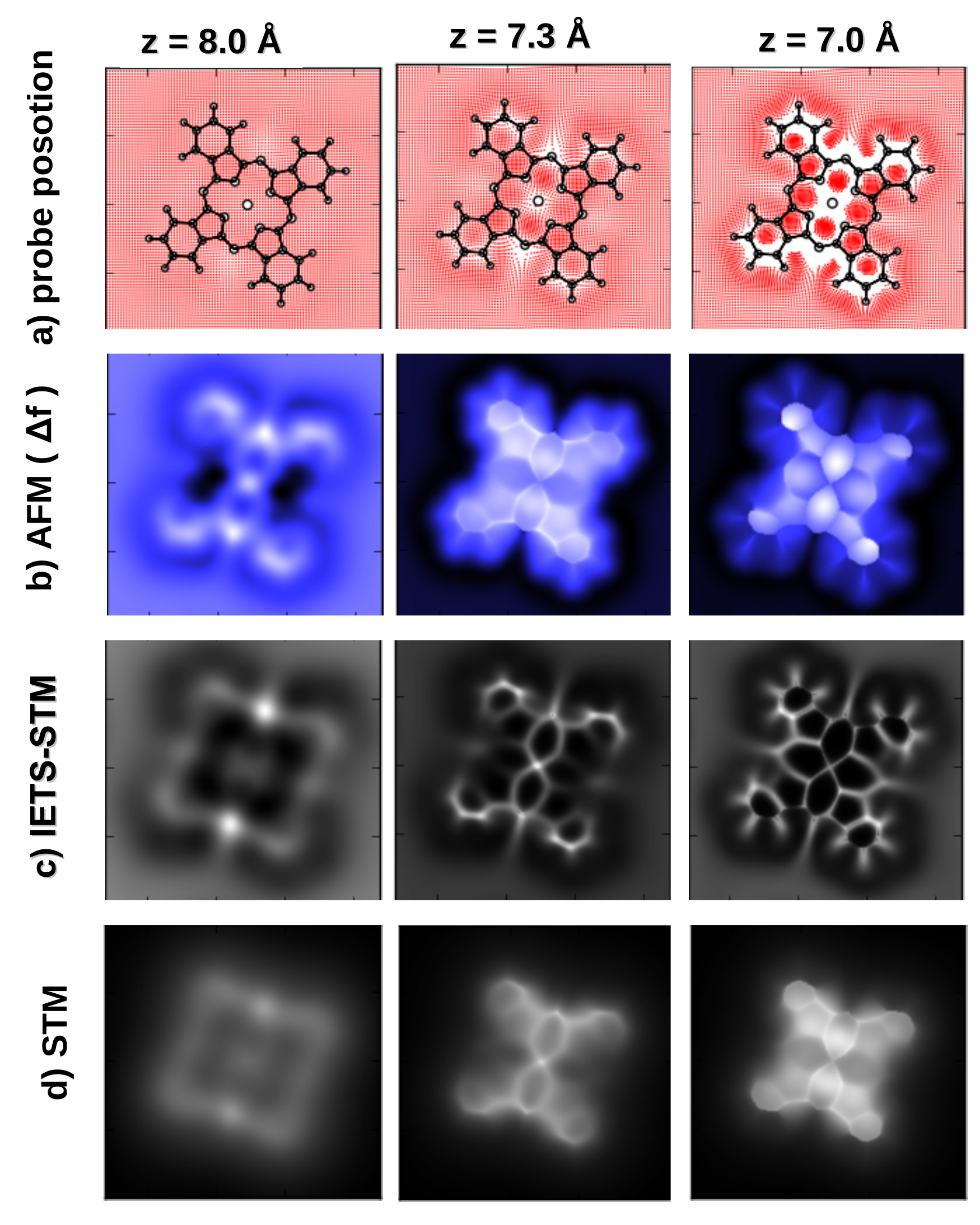}
\caption{\textbf{Simulated constant-height AFM, STM and IETS-STM images} obtained with the extended numerical model assuming the \PP\ charge $-0.4e$. a) probe position after the relaxation b) AFM c) IETS d) STM channels for different distances $z$ = 8.0, 7.3 and 7.0 \AA . Color scale in all images except (a) is renormalized to obtain the best molecular contrast. }
\label{fig-04}
\end{figure}

Finally we compare the different channels (AFM, STM and IETS) calculated with the numerical model, which includes the electrostatic distortion correction considering the probe particle with the negative charge -0.4e.  \reffig{fig-04}a illustrates the relaxations of the probe particle due to the interaction with the surface, which is responsible for the distortion and sharpening of AFM/STM images as discussed in \cite{Gross2012, Hapala_PRB_2014}. In the far distance regime, where the probe particle does not relax, the molecular contrast is blurred in all channels. The situation changes when the probe particle starts to move towards the local minima of the surface potential producing the sharpening of both STM and AFM contrasts followed by the characteristic signal inversion between the atoms/bonds and the rings as the tip approaches the sample \cite{Hapala_PRB_2014}. Simultaneously with the sharpening and inversion of the STM and AFM contrasts we observe a considerable sharpening of the contrast in the IETS-STM channel which unambiguously points out the importance of the \PP\ relaxation for the increased resolution observed in the experimental IETS-STM images. 

We would like to stress that the pattern of the calculated constant-height IETS-STM image in the far distance $z$= 8 \AA\ shown in \reffig{fig-04}c  coincides very well with the Hartree potential projected in the same $z$-plane (see \reffig{fig-02}). This means that the shape of tip-sample potential energy is determined mostly by the electrostatic interaction. Consequently the IETS-STM images acquired in far distances is negligible maps directly the variation of the electrostatic surface potential at the given $z$ distance.
  
To summarize, we show that the frustrated translational mode of CO molecule placed on the tip apex reacts sensitively to the changes of local curvature of the surface potential. Namely, the attractive (convex) and repulsive (concave) character of the surface potential induces vibration mode hardening and softening, respectively. Detection of the resultant variations of the vibrational energy by the standard means of IETS-STM  allows one to map laterally the curvature of the surface potential. Since the curvature of the surface potential changes strongly in the vicinity of the atoms and interatomic bonds the obtained maps are expected to be closely related to the underlying structure of the scanned surface. Finally we show that the structural resolution obtained by lateral mapping of the IETS-STM signal is further sharpened by the lateral relaxations of the particle decorating the tip that occurs under the influence of the repulsive forces acting on it from the surface. That mechanism puts the AFM, STM and IETS-STM imagining with modified tips on a common ground. In addition, we demonstrate that decorated tips can also be used to image local distributions of electrostatic charge. Eventually this observation may be useful for the development and understanding of imaging techniques yielding better spacial resolution than the traditional AFM-based Kelvin probe force microscopy (KPFM)  \cite{Mohn_NatureNano_2013}. We believe that a detail understanding of the IETS imaging mechanism and the influence of the internal charge distribution on the high-resolution images will contribute to further proliferate SPM techniques.

\section{ acknowledgements }
P.H. and P.J. acknowledge the support by GA\v{C}R, grant no.\ 14-16963J. R.T. thanks the Helmholtz Gemeinschaft for financial support in the framework of a Young Investigator Research Group.


\begin{thebibliography}{27}
\bibitem [{not()}]{note1}%
  \BibitemOpen
  \href@noop {} {}\bibinfo {note} {For simplicity we simulate the IETS signal
  as a gaussian peak in $d^2I/dV^2$ centered around the vibration energy
  $\varepsilon$ with a fixed width $\sigma = 0.7$ meV, which fits well the
  experimental evidence.}\BibitemShut {Stop}%
\end{thebibliography}
\end{document}